\documentclass[aps,twocolumn,nofootinbib,floats,eqsecnum,showpacs]{revtex4-1}
\usepackage{amsmath}

\usepackage{bm} % bold math font
\usepackage{mathrsfs} % for Ralph Smith's formal script font

\newcommand{\secref}[1]{$\S$\ref{#1}}

\newcommand{\eqnref}[1]{(\ref{#1})}

\newcommand{\tabref}[1]{Table~\ref{#1}}

\newcommand{\footref}[1]{Footnote~\ref{#1}}

\def\mbeta{\tilde{\beta}}
\def\malpha{\tilde{\alpha}}
\def\mbalpha{\tilde{\boldsymbol{\alpha}}}

\def\mbsigma{\tilde{\boldsymbol{\sigma}}}
\def\mgamma{\tilde{\gamma}}

\begin{document}

\title
{Correspondence between classical and Dirac-Pauli spinors in view of the Foldy-Wouthuysen transformation}

\author{Tsung-Wei Chen}
\email{twchen@mail.nsysu.edu.tw} \affiliation{Department of Physics, National Sun Yat-sen University, Kaohsiung 80424, Taiwan}

\author{Dah-Wei Chiou}
\email{dwchiou@gmail.com}
\affiliation{Department of Physics and Center for Condensed Matter Sciences, National Taiwan University, Taipei 10617, Taiwan}

\begin{abstract}
The classical dynamics for a charged spin particle is governed by the Lorentz force equation for orbital motion and by the Thomas-Bargmann-Michel-Telegdi (T-BMT) equation for spin precession. In static and homogeneous electromagnetic fields, it has been shown that the Foldy-Wouthuysen (FW) transform of the Dirac-Pauli Hamiltonian, which describes the relativistic quantum theory for a spin-$1/2$ particle, is consistent with the classical Hamiltonian (with both the orbital and spin parts) up to the order of $1/m^{14}$ ($m$ is the particle's mass) in the low-energy/weak-field limit. In this paper, we extend this correspondence to the case of inhomogeneous fields. Regardless of the field gradient (e.g., Stern-Gerlach) force, the T-BMT equation is unaltered and thus the classical Hamiltonian remains the same, but subtleties arise and need to be clarified. For the relativistic quantum theory, we apply Eriksen's method to obtain the exact FW transformations for the two special cases, which in conjunction strongly suggest that, in the weak-field limit, the FW transformed Dirac-Pauli Hamiltonian (except for the Darwin term) is in agreement with the classical Hamiltonian in a manner that classical variables correspond to quantum operators via a specific Weyl ordering. Meanwhile, the Darwin term is shown to have no classical correspondence.
\end{abstract}


\maketitle

\section{Introduction}\label{sec:introduction}
The relativistic quantum theory for a spin-$1/2$ point particle is described by the Dirac equation \cite{Dirac1928,Dirac1982}. The wavefunction used for the Dirac equation is the Dirac bispinor, which is composed of two Weyl spinors corresponding to the particle and antiparticle parts. Rigorously, the Dirac equation is self-consistent only in the context of quantum field theory, in which the particle-antiparticle pairs can be created and annihilated. In the low-energy limit, if the relevant energy (the particle's energy interacting with electromagnetic fields) is much smaller than the Dirac energy gap $2mc^2$ ($m$ is the particle's mass), the
probability of particle-antiparticle pair creation/annihilation is negligible and the Dirac equation, after block diagonalization, is adequate to describe the relativistic quantum dynamics of the spin-$1/2$ particle without taking into account the field-theory interaction to the antiparticle.

The Foldy-Wouthuysen (FW) transformation is one of the methods used to investigate the low-energy limit of the Dirac equation via a series of successive unitary transformations, each of which block-diagonalizes the Dirac Hamiltonian to a certain order of $1/m$ \cite{Foldy1950} (see \cite{Strange2008} for a review). Following the standard FW method, many different approaches have been developed for various advantages \cite{Lowding1951,Luttinger1955,Eriksen1958,Kutzelnigg1990,Winkler2003}.
Particularly, Kutzelnigg proposed a self-consistent equation which allows one to obtain the block-diagonalized Dirac Hamiltonian without evoking decomposition of even and odd Dirac matrices \cite{Kutzelnigg1990}.
For the spinor subject to nonexplicitly time-dependent fields, Eriksen developed a systematic method of a single unitary transformation, which gives an exact FW transformation when the interaction term is an odd Dirac matrix \cite{Eriksen1958}. The validity of Eriksen's method was investigated in \cite{Vries1968}.

Alternatively, the Dirac Hamiltonian can also be diagonalized via the method of expansion in powers of the Planck constant $\hbar$ \cite{Silenko2003,Bliokh2005,Goss2007}, which enables one to investigate the quantum corrections on the classical dynamics in strong fields \cite{Silenko2008}.

For the case that the gyromagnetic ratio is different from $e/(mc)$, the relativistic quantum theory of the spin-$1/2$ particle with inclusion of the anomalous magnetic moment can be described by the Dirac-Pauli equation, which is the Dirac equation augmented with extra terms explicitly dependent on electromagnetic field strength \cite{Pauli1941}. The FW transformation method for the Dirac Hamiltonian can be straightforwardly applied to the Dirac-Pauli Hamiltonian \cite{Silenko2008}.

On the other hand, the classical (non-quantum) dynamics for a relativistic point particle endowed with charge and intrinsic spin in static and homogeneous electromagnetic fields is well understood. The orbital motion is governed by the Lorentz force equation and the precession of spin by the Thomas-Bargmann-Michel-Telegdi (T-BMT) equation \cite{Thomas1927,Bargmann1959} (see $\S$11 of \cite{Jackson1999} for a review). The orbital Hamiltonian for the Lorentz force equation plus the spin Hamiltonian for the T-BMT equation is expected to provide a low-energy description of the relativistic quantum theory. It is natural to conjecture that, in the low-energy/weak-field limit, the Dirac or, more generically, Dirac-Pauli Hamiltonian reduces to the sum of the classical orbital and spin Hamiltonians.

The correspondence between the Dirac-Pauli Hamiltonian and the classical Hamiltonian has been investigated from various aspects with or without external electromagnetic fields. For a free Dirac spinor, it has been shown that the exact FW transformed Dirac Hamiltonian corresponds to the relativistic orbital Hamiltonian \cite{Foldy1950}. In \cite{Rubinow1963,Rafa1964}, it was shown that the T-BMT equation may be derived from the WKB solution to the wavefunction of the Dirac equation. In the presence of external electromagnetic fields, the FW transformation on the Dirac Hamiltonian has been performed up to the order of $1/m^4$, but the connection to the classical counterpart was not explicit \cite{Froh1993}; the explicit connection was addressed in \cite{Silenko1995}. In \cite{Chen2010}, it has been shown that, in static and homogeneous electromagnetic fields, the resulting FW transformed Dirac-Pauli Hamiltonian is in full agreement with the classical Hamiltonian up to the order of $1/m^8$, if nonlinear terms of field strength are neglected in the weak-field limit. Recently, the work of \cite{Chen2010} was extended to the order of $1/m^{14}$ by applying Kutzelnigg's method \cite{Chang:thesis}; Kutzelnigg's method can be further simplified so that the FW transformation can be obtained more systematically and efficiently to any desired order \cite{paper-to-appear}.

In this paper, we try to extend the correspondence between classical and Dirac-Pauli spinors as affirmed in \cite{Chen2010,Chang:thesis,paper-to-appear} to the more generic case of inhomogeneous electromagnetic fields. Complications arise even at the level of classical dynamics. In addition to the Lorentz force, the orbital motion also experiences the force due to field strength gradient such as the Stern-Gerlach force. This in turn modifies the BMT equation (covariant counterpart of the T-BMT equation). We derive the modified BMT equation and prove that, regardless of the modification, the corresponding T-BMT equation remains the same. Furthermore, when the orbital Hamiltonian and spin Hamiltonian are added together, subtleties of the canonical formalism call forth more careful investigation on the physical interpretations of canonical variables.

For the relativistic quantum theory, on the other hand, complications mainly come from issues of operator ordering, as kinematic momentum operators in three directions commute neither with one another nor with electromagnetic fields due to inhomogeneity of fields. As a result, it is very cumbersome to obtain FW transformations in an order-by-order scenario. Instead, we consider two special cases of which the interaction term is an odd Dirac matrix, and Eriksen's method is used to obtain the exact FW transformed Hamiltonian for both cases. These two special cases joined together strongly suggest that, in the weak-field limit, the FW transformed Dirac-Pauli Hamiltonian (except for the Darwin term) agrees with the classical Hamiltonian in a manner that classical variables correspond to quantum operators via a specific Weyl ordering. Additionally, the Darwin term is shown to have no direct classical counterpart.

This paper is organized as two parts. In \secref{sec:classical relativistic spinor}, we study the classical relativistic spinor, deriving the (T-)BMT equations and investigating the Hamiltonian formalism. In \secref{sec:Dirac-Pauli spinor}, we study the Dirac-Pauli spinor, showing that the FW transformed Dirac-Pauli Hamiltonian obtained by Eriksen's method agrees with the classical counterpart plus the Darwin term. Conclusions are summarized and discussed in \secref{sec:summary}.

\section{Classical relativistic spinor}\label{sec:classical relativistic spinor}
As the first part, we study the classical relativistic spinor. In the presence of inhomogeneity of electromagnetic fields, the orbital motion experiences the field gradient (e.g., Stern-Gerlach) force in addition to the familiar Lorentz force. This in turn modifies the BMT equation (covariant counterpart of the T-BMT equation) for the spin precession, whereas the T-BMT equation is shown to remain unaltered. We scrutinize the Hamiltonian formalism and clarity subtleties of the physical interpretations of canonical variables.

\subsection{Spin 4-vector}\label{sec:spin 4-vector}
A particle endowed with intrinsic spin (call ``spinor'') can give rise to an \emph{intrinsic} magnetic moment whether it is charged (e.g., electron) or not (e.g., neutron). The fact that magnetic dipole moment $\boldsymbol{\mu}_m$ and electric dipole moment $\boldsymbol{\mu}_e$ form an antisymmetric tensor $M^{\mu\nu}$ suggests that the intrinsic spin $\mathbf{s}$ can be generalized to a second-rank antisymmetric tensor $S^{\mu\nu}$, which gives the intrinsic dipole moments as
\begin{equation}\label{M and S}
M^{\mu\nu}=\gamma_m S^{\mu\nu},
\end{equation}
where $\gamma_m$ is the gyromagnetic ratio.\footnote{\label{foot:gyromagnetic ratio}The gyromagnetic ratio is usually given as $\gamma_m=\frac{ge}{2mc}$, where $g$ is the $g$-factor. We deliberately keep $\gamma_m$ independent of $e$ and $m$ in order to take into account the case which is of zero charge ($e=0$) but nevertheless with nonzero magnetic moment ($\gamma_m\neq0$). For convenience, we also define $\mu:=\gamma_m\hbar/2$ for later use.} The spin has only three independent components; thus $S^{\mu\nu}$ is dual to an axial 4-vector $S^\alpha=(S^0,\mathbf{S})$ via
\begin{equation}\label{S tensor}
S^{\mu\nu}=\frac{1}{c}\,\epsilon^{\mu\nu\alpha\beta}U_\alpha S_\beta
\end{equation}
and conversely
\begin{equation}
S^\alpha=\frac{1}{2c}\,\epsilon^{\alpha\beta\gamma\delta}U_\beta S_{\gamma\delta},
\end{equation}
where $U^\alpha$ is the particle's 4-velocity. The 4-vector $S^\alpha$ in the laboratory (unprimed) frame reduces to the spin $\mathbf{s}$ in the particle's rest (primed) frame; i.e., $S'^\alpha=(S'^0,\mathbf{S}')\equiv(0,\mathbf{s})$.\footnote{And accordingly,
\begin{equation}\label{S' matrix}
S^{\prime\alpha\beta}=\left(
                         \begin{array}{cccc}
                           0 & 0 & 0 & 0 \\
                           0 & 0 & -s_z & s_y \\
                           0 & s_z & 0 & -s_x \\
                           0 & -s_y & s_x & 0 \\
                         \end{array}
                       \right)
   =S'_{\alpha\beta}.
\end{equation}} The vanishing of the time component in the particle's rest frame is imposed by the covariant constraint
\begin{equation}\label{constraint 1}
U_\alpha S^\alpha=0,
\end{equation}
or equivalently
\begin{equation}\label{constraint 2}
S^{\alpha\beta}U_\beta=0.
\end{equation}
In the particle's rest frame, $U^{\prime\alpha}=(c,\mathbf{0})$ and \eqnref{M and S} yields
\begin{equation}
\boldsymbol{\mu}'_m=\gamma_m\mathbf{s},
\qquad
\boldsymbol{\mu}'_p=0.
\end{equation}

\subsection{Thomas-Bargmann-Michel-Telegdi equations}\label{sec:T-BMT equations}
Consider a relativistic point particle endowed with electric charge and intrinsic spin subject to external electromagnetic fields. In covariant expression, the orbital motion of the particle is described by
\begin{equation}\label{covariant Lorentz eq}
m\frac{dU^\alpha}{d\tau}=\frac{e}{c}\,F^{\alpha\beta}U_\beta +f^\alpha,
\end{equation}
where the first term on the right-hand side is the familiar Lorentz force and the second term $f^\alpha$ is included on account of any other forces such as the field gradient force or even nonelectromagnetic forces.

On the other hand, exploiting Lorentz covariance, we infer that the equation of spin precession must be of the covariant form (see \cite{Bargmann1959} and also $\S$11.11 of \cite{Jackson1999})
\begin{eqnarray}\label{dS dtau}
\frac{dS^\alpha}{d\tau} &=& A_1F^{\alpha\beta}S_\beta + \frac{A_2}{c^2}(S_\lambda F^{\lambda\mu}U_\mu)U^\alpha\nonumber\\
&&\mbox{}+ \frac{A_3}{c^2}\left(S_\beta\frac{dU^\beta}{d\tau}\right)U^\alpha,
\end{eqnarray}
where $A_i$ are constants and we have assumed that the equation is linear in the spin $S^\alpha$ and in the external fields $F^{\alpha\beta}$ and higher time derivatives are absent.
The constraint \eqnref{constraint 1} must hold at all times, which requires
\begin{equation}\label{dUS dtau}
\frac{d}{d\tau}(U_\alpha S^\alpha) = S^\alpha\frac{dU_\alpha}{d\tau} +U_\alpha\frac{dS^\alpha}{d\tau} =0.
\end{equation}
By \eqnref{constraint 1}, \eqnref{dS dtau}, and $U^\alpha U_\alpha=c^2$, \eqnref{dUS dtau} then gives
\begin{equation}
(A_1-A_2)U_\alpha F^{\alpha\beta}S_\beta + (1+A_3)S_\beta\frac{dU^\beta}{d\tau} =0
\end{equation}
for arbitrary $F^{\alpha\beta}$. This follows $A_1=A_2$ and $A_3=-1$. Moreover, reduction to the rest frame from \eqnref{dS dtau} (with $S'^\alpha=(0,\mathbf{s})$ and $U'^\alpha=(c,\mathbf{0})$) yields
\begin{equation}
\frac{d\mathbf{s}}{dt'}=A_1\mathbf{s}\times\mathbf{B}',
\end{equation}
which, to conform with the familiar Larmor precession, fixes $A_1=\gamma_m$. Consequently, \eqnref{dS dtau} gives the BMT equation (modified with the possible presence of $f^\alpha$)
\begin{eqnarray}\label{BMT}
\frac{dS^\alpha}{d\tau}&=&
\gamma_mF^{\alpha\beta}S_\beta
+\frac{1}{c^2}\left(\gamma_m-\frac{e}{mc}\right)U^\alpha
\left(S_\lambda F^{\lambda\mu} U_\mu\right)\nonumber\\
&&\mbox{}-\frac{1}{mc^2}U^\alpha S_\lambda f^\lambda.
\end{eqnarray}

In order to obtain $dS'^{\mu}/dt$, we use the Lorentz transformation $S^{\mu}={\Lambda_{\lambda}}^\mu S'^{\lambda}$ and have
$\frac{dS^{\mu}}{dx^0}= \frac{d{\Lambda_\lambda}^\mu}{dx^0}S'^{\lambda} +{\Lambda_\lambda}^\mu\frac{dS'^{\lambda}}{dx^0}$.
Contracting index $\mu$ by ${\Lambda^{\rho}}_\mu$, we obtain
\begin{equation}\label{T-BMT1}
\frac{dS'^{\rho}}{dx^0}
=-{\Lambda^{\rho}}_\mu\frac{d{\Lambda_\lambda}^\mu}{dx^0}S'^{\lambda}
+\frac{1}{U^0}\Lambda^{\rho}_{~\mu}\frac{dS^{\mu}}{d\tau},
\end{equation}
where
${\Lambda^{\rho}}_\mu{\Lambda_{\lambda}}^\mu=g^{\rho}_{\lambda}$ and $U^{\mu}=dx^{\mu}/d\tau$ have been used. By \eqnref{BMT}, the second term of \eqnref{T-BMT1} can be written as
\begin{eqnarray}\label{T-BMT2}
{\Lambda^{\rho}}_\mu\frac{dS^{\mu}}{d\tau}
&=&\gamma_m
\left({\Lambda^{\rho}}_\mu F^{\mu\nu}{\Lambda^{\sigma}}_\nu\right)S'_{\sigma}\nonumber\\
&&\mbox{}
-\frac{1}{c^2}\left(\gamma_m-\frac{e}{mc}\right)
U_{\alpha}F^{\alpha\beta}{\Lambda^{\lambda}}_\beta S'_{\lambda}U'^{\rho}\nonumber\\
&&\mbox{}
+\frac{1}{mc^2}f^\beta{\Lambda^{\lambda}}_\beta S'_{\lambda}U'^{\rho},
\end{eqnarray}
where $S_{\nu}={\Lambda^{\sigma}}_\nu S'_{\sigma}$ and
$S_{\beta}={\Lambda^{\lambda}}_\beta S'_{\lambda}$ have been used. In the particle's rest frame, both the spatial component of the 4-velocity and time component of spin are zero (i.e., $U'^i=0$ and $S'^0=0$). Therefore,
by taking \eqnref{T-BMT2} into \eqnref{T-BMT1}, the spatial component of $\frac{dS'^{\rho}}{dx^0}$ reads as
\begin{equation}\label{T-BMT3}
\frac{dS'^i}{dt}=-{\Lambda^i}_\mu \frac{d{\Lambda_j}^\mu}{dt}S'^j
+\gamma_m\frac{c}{U_0}\left({\Lambda^i}_\mu
F^{\mu\nu}{\Lambda^k}_\nu\right)S'_k,
\end{equation}
where the Lorentz transformation matrix is given by (see $\S$11.7 of \cite{Jackson1999})
\begin{subequations}\label{Lambda}
\begin{eqnarray}
{\Lambda^0}_0&=&\gamma,\\
{\Lambda^i}_0&=&-\gamma\beta^i\equiv-\frac{U^i}{c},\\
{\Lambda^i}_j&=&g^i_j+\frac{(\gamma-1)\beta^i\beta^j}{\beta^2}
\equiv g^i_j-\frac{1}{1+\gamma}\frac{U^iU_j}{c^2},\qquad
\end{eqnarray}
\end{subequations}
the Lorentz factor $\gamma$ is given by
\begin{equation}\label{gamma}
\gamma:=\frac{1}{\sqrt{1-\boldsymbol{\beta}^2}},
\qquad
\boldsymbol{\beta}:=\frac{\mathbf{v}}{c}
\end{equation}
with $\mathbf{v}\equiv d\mathbf{x}/dt$ being the particle's (boost) velocity,
and the electric field $\mathbf{E}$ and magnetic field $\mathbf{B}$ are given via
\begin{equation}\label{F}
F^{0i}=-E^i,
\qquad F^{ij}=-\epsilon_{ijk}B^k.
\end{equation}
We can express \eqnref{T-BMT3} in terms of $\mathbf{E}$, $\mathbf{B}$, $\gamma$, and $\boldsymbol{\beta}$.
By \eqnref{Lambda} and \eqnref{F}, after tedious but straightforward calculations, it can be shown that
\begin{subequations}\label{T-BMT4}
\begin{eqnarray}
&&\Lambda^i_{~\mu}\frac{d\Lambda^{~\mu}_{j}}{dt}S'^j\\
&=&-\frac{e}{mc}\frac{\gamma}{\gamma+1}
\left\{\mathbf{S}'\times\left[\boldsymbol{\beta}\times\mathbf{E}
+\boldsymbol{\beta}\times\left(\boldsymbol{\beta}\times\mathbf{B}\right)\right]\right\}^i,\nonumber\\
&&\left(\Lambda^i_{~\mu}F^{\mu\nu}\Lambda^k_{~\nu}\right)S'_k\\
&=&\gamma\left\{\mathbf{S}'\times\left[\mathbf{E}\times\boldsymbol{\beta}
+\frac{1}{\gamma}\mathbf{B}
-\frac{\gamma}{1+\gamma}\boldsymbol{\beta}\times
\left(\boldsymbol{\beta}\times\mathbf{B}\right)\right]\right\}^i.\nonumber
\end{eqnarray}
\end{subequations}
Equation~\eqnref{T-BMT3} with substitution of \eqnref{T-BMT4} can be written, after simplification, as
\begin{equation}\label{T-BMT5}
\frac{d\mathbf{s}}{dt}=\mathbf{s}\times\mathbf{F}
\end{equation}
with
\begin{eqnarray}\label{T-BMT F}
\mathbf{F}&=&\left(\gamma_m-\frac{e}{mc}+\frac{e}{mc}\frac{1}{\gamma}\right)\mathbf{B}\nonumber\\
&&\mbox{}
-\left(\gamma_m-\frac{e}{mc}\right)\frac{\gamma}{\gamma+1}
\left(\boldsymbol{\beta}\cdot\mathbf{B}\right)\boldsymbol{\beta}\nonumber\\
&&\mbox{}
-\left(\gamma_m-\frac{e}{mc}\frac{\gamma}{\gamma+1}\right)\boldsymbol{\beta}\times\mathbf{E},
\end{eqnarray}
where we have used $S'^\alpha=(S'^0,\mathbf{S}')\equiv(0,\mathbf{s})$. Equation \eqnref{T-BMT5} with \eqnref{T-BMT F} is called the T-BMT equation.\footnote{We call \eqnref{BMT} the (modified) BMT equation and \eqnref{T-BMT5} the T-BMT equation. In the literature, \eqnref{BMT} and \eqnref{T-BMT5} are also referred to as the first and second T-BMT equation respectively.} Note that \eqnref{T-BMT5} immediately implies that the magnitude of $\mathbf{s}$ is conserved; i.e.,
\begin{equation}
\frac{d\mathbf{s}^2}{dt}=0.
\end{equation}
Also note that inclusion of $f^\alpha$ in \eqnref{covariant Lorentz eq} modifies the BMT equation in \eqnref{BMT}, but the T-BMT equation \eqnref{T-BMT5} remains the same whether $f^\alpha$ is taken into account or not.\footnote{That is, while the orbital motion $d\mathbf{p}/dt$ is sensitive to inhomogeneity of external fields, the spin precession $d\mathbf{s}/dt$ is sensitive only to the field strength but not its inhomogeneity. The effect of $f^\alpha$ on $dS^\alpha/d\tau$ is indirectly through the Lorentz transformation.}

In the low-speed limit ($\beta\ll 1$), we have $\gamma\approx1$
and \eqnref{T-BMT5} with \eqnref{T-BMT F} yields
\begin{eqnarray}\label{T-BMT low speed}
\frac{d\mathbf{s}}{dt}
&\approx&\mathbf{s}\times
\left[
\gamma_m\mathbf{B}
-\frac{1}{2}\left(\gamma_m-\frac{e}{mc}\right)(\boldsymbol{\beta}\cdot\mathbf{B})\boldsymbol{\beta}
\right.\nonumber\\
&&\qquad\quad
\left.
-\left(\gamma_m-\frac{e}{2mc}\right)\boldsymbol{\beta}\times\mathbf{E}
\right].
\end{eqnarray}
The first term in \eqnref{T-BMT low speed} accounts for the torque acting on the magnetic moment $\boldsymbol{\mu}_m$ by the magnetic field. The second term corresponds to the change rate of the longitudinal polarization, which vanishes in the case of $g=2$. The third term is the spin-orbit interaction (the interaction of the boosted electric dipole $\boldsymbol{\mu}_p\approx\boldsymbol{\beta}\times\boldsymbol{\mu}'_m$ coupled to the electric field) plus the correction due to the Thomas precession \cite{Thomas1927}.

\subsection{Hamiltonian formalism}\label{sec:Hamiltonian formalism}
In the laboratory frame, the orbital motion of a charged particle is governed by the Hamiltonian $H_\mathrm{orbit}$:
\begin{equation}\label{H orbit}
H_\mathrm{orbit}(\mathbf{x},\mathbf{p};t)
=\sqrt{c^2\boldsymbol{\pi}^2+m^2c^4}\,
+e\,\phi(\mathbf{x},t),
\end{equation}
where the kinematic momentum $\pi$ is defined as
\begin{equation}\label{def pi}
\boldsymbol{\pi}:=\mathbf{p}-\frac{e}{c}\,\mathbf{A}(\mathbf{x},t),
\end{equation}
$\mathbf{p}$ is the canonical momentum conjugate to $\mathbf{x}$ with the canonical relation $\{x_i,p_j\}=\delta_{ij}$, and $A^\alpha=(\phi,\mathbf{A})$ is the 4-potential for electromagnetic fields.
Hamilton's equations are
\begin{subequations}
\begin{eqnarray}
\frac{d\mathbf{x}}{dt}&=&\{\mathbf{x},H_\mathrm{orbit}\}=\boldsymbol{\nabla}_\mathbf{p}H_\mathrm{orbit},\\
\frac{d\mathbf{p}}{dt}&=&\{\mathbf{p},H_\mathrm{orbit}\}=-\boldsymbol{\nabla} H_\mathrm{orbit},
\end{eqnarray}
\end{subequations}
which lead to
\begin{equation}\label{orbit eq 1}
\mathbf{v}\equiv\frac{d\mathbf{x}}{dt}
=\frac{\boldsymbol{\pi}}{m\gamma_{\boldsymbol{\pi}}}
\end{equation}
and\footnote{The identity $\boldsymbol{\nabla}(\mathbf{a}\cdot\mathbf{b}) =(\mathbf{a}\cdot\boldsymbol{\nabla})\mathbf{b} +(\mathbf{b}\cdot\boldsymbol{\nabla})\mathbf{a} +\mathbf{a}\times(\boldsymbol{\nabla}\times\mathbf{b}) +\mathbf{b}\times(\boldsymbol{\nabla}\times\mathbf{a})$ is used.}
\begin{equation}\label{orbit eq 2}
\frac{d\mathbf{p}}{dt} = \frac{e}{mc\gamma_{\boldsymbol{\pi}}}
\left[(\boldsymbol{\pi}\cdot\boldsymbol{\nabla})\mathbf{A}
+\boldsymbol{\pi}\times(\boldsymbol{\nabla}\times\mathbf{A})\right]
-e\boldsymbol{\nabla}\phi,
\end{equation}
respectively, where the Lorentz factor associated with $\boldsymbol{\pi}$ is defined as
\begin{equation}\label{gamma pi}
\gamma_{\boldsymbol{\pi}}:=\sqrt{1+\left(\frac{\boldsymbol{\pi}}{mc}\right)^2}\,.
\end{equation}
Comparing \eqnref{gamma}, \eqnref{orbit eq 1}, and \eqnref{gamma pi}, we find
\begin{equation}\label{pi and v}
\gamma_{\boldsymbol{\pi}}=\gamma,
\qquad
\boldsymbol{\pi}=\gamma m\mathbf{v}\equiv m\mathbf{U}.
\end{equation}
With \eqnref{pi and v} and by the identity
${d\mathbf{A}}/{dt}={\partial\mathbf{A}}/{\partial t}+(\mathbf{v}\cdot\boldsymbol{\nabla})\mathbf{A}$,
\eqnref{orbit eq 2} leads to the familiar Lorentz equation
\begin{eqnarray}\label{Lorentz eq}
&&\frac{d\boldsymbol{\pi}}{dt}\equiv\frac{d\left(\mathbf{p}-\frac{e}{c}\mathbf{A}\right)}{dt}\nonumber\\
&=&
-e\left(\boldsymbol{\nabla}\phi-\frac{1}{c}\frac{\partial\mathbf{A}}{\partial t}\right)
+e\,\frac{\mathbf{v}}{c}\times(\boldsymbol{\nabla}\times\mathbf{A})\nonumber\\
&=&e\left(\mathbf{E}+\frac{\mathbf{v}}{c}\times\mathbf{B}\right),
\end{eqnarray}
which can be shown to be equivalent to \eqnref{covariant Lorentz eq} (with $f^\alpha=0$) by noting that $U^\alpha=(\gamma c,\gamma\mathbf{v})$ and $\gamma d\tau=dt$.

On the other hand, the equation of spin precession \eqnref{T-BMT5} can be recast as Hamilton's equation
\begin{equation}\label{ds dt}
\frac{d\mathbf{s}}{dt}=\{\mathbf{s},H_\mathrm{spin}\}
\end{equation}
subject to the canonical relation
\begin{equation}
\{s_i,s_j\}=\epsilon_{ijk}s_k,
\end{equation}
as the Hamiltonian for the spin precession is given by
\begin{equation}\label{H spin}
H_\mathrm{spin}(\mathbf{s})=-\mathbf{s}\cdot \mathbf{F}(\mathbf{x},t)
\end{equation}
with the same $\mathbf{F}$ as defined in \eqnref{T-BMT F}.

As the orbital motion and the spin precession admit Hamiltonian formalisms in the laboratory frame separately, it is natural to expect that the sum of the two Hamiltonians, i.e.,
\begin{equation}\label{H total}
H=H_\mathrm{orbit}+H_\mathrm{spin},
\end{equation}
describes both the orbital motion and the spin precession simultaneously, if we treat $\mathbf{s}$ as phase-space variables independent of $\mathbf{x}$ and $\mathbf{p}$.\footnote{Note that, in order to put $H_\mathrm{orbit}$ and $H_\mathrm{spin}$ on the same footing, we have to consider $d\mathbf{s}/dt\equiv d\mathbf{S}'/dt$ in \eqnref{T-BMT5}, instead of $d\mathbf{S}/dt$, $d\mathbf{S}/d\tau$, or $d\mathbf{s}/d\tau$. This is because $s_i$ are degrees of freedom independent of $\mathbf{x}$ and $\mathbf{p}$, but $S_i$ are not. Furthermore, in accord with the orbital motion, the precession is cast with respect to $t$, instead of the proper time $\tau$ of the moving particle.}
Indeed, we have
\begin{equation}
\frac{d\mathbf{p}}{dt}=-\boldsymbol{\nabla} H_\mathrm{orbit}-\boldsymbol{\nabla} H_\mathrm{spin},
\end{equation}
which yields the desired result as the first term on the right-hand side gives the Lorentz force and the second term gives the Stern-Gerlach force due to field gradient.
On the other hand, however, because $\boldsymbol{\beta}$ and $\gamma$ in \eqnref{T-BMT F} involve $\mathbf{p}$ (through $\boldsymbol{\pi}$), $\boldsymbol{\nabla}_\mathbf{p}$ will also hit $H_\mathrm{spin}$ and consequently we no longer have the simple relation \eqnref{orbit eq 1} but instead
\begin{eqnarray}
\mathbf{v}\equiv\frac{d\mathbf{x}}{dt}
&=&\boldsymbol{\nabla}_\mathbf{p}H_\mathrm{orbit}+\boldsymbol{\nabla}_\mathbf{p}H_\mathrm{spin}\nonumber\\
&=&\frac{\boldsymbol{\pi}}{m\gamma_{\boldsymbol{\pi}}}+\boldsymbol{\nabla}_\mathbf{p}H_\mathrm{spin},
\end{eqnarray}
which implies that $\mathbf{v}$ not only involves $\mathbf{p}$ but also $\mathbf{s}$, although the involvement with $\mathbf{s}$ is negligible if $H_\mathrm{spin}\ll mc^2$ (which is true in the weak-field limit). This in turn makes $\boldsymbol{\beta}$ and $\gamma$ become (complicated) functions of both $\mathbf{p}$ and $\mathbf{s}$ (and particularly $\gamma\neq\gamma_{\boldsymbol{\pi}}$). As a result, \eqnref{ds dt} no longer yields the desired T-BMT equation \eqnref{T-BMT5}.

Comparison to the FW transform of the Dirac-Pauli Hamiltonian (see previous works \cite{Chen2010,Chang:thesis,paper-to-appear}), however, suggests that the aforementioned complication can be avoided by simply replacing $\gamma$ and $\boldsymbol{\beta}$ in \eqnref{T-BMT F} with their counterparts associated with $\boldsymbol{\pi}$. That is, defining the 4-``velocity'' associated with $\boldsymbol{\pi}$ as
\begin{equation}\label{U pi}
U^\alpha_{\boldsymbol{\pi}}:=\left(\gamma_{\boldsymbol{\pi}}c,\frac{\boldsymbol{\pi}}{m}\right)
\equiv(\gamma_{\boldsymbol{\pi}}c,\gamma_{\boldsymbol{\pi}}\mathbf{v}_{\boldsymbol{\pi}}),
\end{equation}
which gives $U_{\boldsymbol{\pi}}^\alpha U_{{\boldsymbol{\pi}}\alpha}=c^2$, and accordingly the ``velocity'' associated with $\boldsymbol{\pi}$ as
\begin{equation}\label{v pi}
c\,\boldsymbol{\beta}_{\boldsymbol{\pi}}\equiv\mathbf{v}_{\boldsymbol{\pi}}:=
\frac{\boldsymbol{\pi}}{\gamma_{\boldsymbol{\pi}}m},
\end{equation}
$H_\mathrm{spin}$ in \eqnref{H spin} is to be modified by replacing every $\gamma$ and $\boldsymbol{\beta}$ with $\gamma_{\boldsymbol{\pi}}$ and $\boldsymbol{\beta}_{\boldsymbol{\pi}}$ respectively in \eqnref{T-BMT F}.\footnote{This prescription was taken for granted in \cite{Chen2010,Chang:thesis} when compared to the FW transformation for the Dirac-Pauli equation; the subtle complications where not explicitly noted.}
To sum up, for describing the orbital motion and the spin precession simultaneously with $\mathbf{x}$, $\mathbf{p}$, and $\mathbf{s}$ as independent phase-space variables, \eqnref{H total} is not completely correct but more accurately we should have the total Hamiltonian as
\begin{eqnarray}\label{H total modified}
&&H(\mathbf{x},\mathbf{p},\mathbf{s};t)\\
&=&H_\mathrm{orbit}(\mathbf{x},\mathbf{p};t)
+H_\mathrm{spin}(\mathbf{s},\mathbf{x},\mathbf{p};t)+\mathcal{O}(F_{\mu\nu}^2,\hbar^2),\nonumber
\end{eqnarray}
where the orbital Hamiltonian $H_\mathrm{orbit}(\mathbf{x},\mathbf{p};t)$ is given by \eqnref{H orbit}, the new spin Hamiltonian $H_\mathrm{spin}(\mathbf{s},\mathbf{x},\mathbf{p};t)$ is given by
\begin{equation}\label{H spin modified}
H_\mathrm{spin}(\mathbf{s},\mathbf{x},\mathbf{p};t)
=-\mathbf{s}\cdot\mathbf{F}_{\boldsymbol{\pi}}(\mathbf{x},\mathbf{p},t)
\end{equation}
with
\begin{eqnarray}\label{T-BMT new F}
\mathbf{F}_{\boldsymbol{\pi}}(\mathbf{x},\mathbf{p},t)&=&
\left(\gamma_m-\frac{e}{mc}
+\frac{e}{mc}\frac{1}{\gamma_{\boldsymbol{\pi}}}\right)\mathbf{B}(\mathbf{x},t)\\
&&\mbox{}
-\left(\gamma_m-\frac{e}{mc}\right)
\frac{\left(\boldsymbol{\pi}\cdot\mathbf{B}(\mathbf{x},t)\right)\boldsymbol{\pi}}
{\gamma_{\boldsymbol{\pi}}(\gamma_{\boldsymbol{\pi}}+1)m^2c^2}\nonumber\\
&&\mbox{}
-\left(\gamma_m-\frac{e}{mc}
\frac{\gamma_{\boldsymbol{\pi}}}{\gamma_{\boldsymbol{\pi}}+1}\right)
\frac{\boldsymbol{\pi}\times\mathbf{E}(\mathbf{x},t)}{\gamma_{\boldsymbol{\pi}}mc},
\nonumber
\end{eqnarray}
and $\mathcal{O}(F_{\mu\nu}^2,\hbar^2)$ denotes other contributions which are negligible in the low-weak limit and/or in the classical (non-quantum) limit as will be discussed shortly.\footnote{The classical limit corresponds to keeping terms up to $\mathcal{O}(\hbar^0)$ for the orbital Hamiltonian and to $\mathcal{O}(\hbar)$ for the spin Hamiltonian, since $\mathbf{s}=\hbar\boldsymbol{\sigma}/2$ in comparison to the quantum theory.}

The simple prescription, although approved of by the previous works \cite{Chen2010,Chang:thesis,paper-to-appear} (up to very high orders for the case of static and homogeneous fields), calls into question whether the new T-BMT equation of spin precession, i.e.,
\begin{equation}\label{TMBT new}
\frac{d\mathbf{s}}{dt}=\mathbf{s}\times\mathbf{F}_{\boldsymbol{\pi}},
\end{equation}
still respects Lorentz covariance as the old T-BMT equation \eqnref{T-BMT5} does (note that \eqnref{T-BMT5} is derived by starting from the Lorentz covariant form in \eqnref{dS dtau}).
The answer turns out to be affirmative, as one can readily reproduce every single formula in \secref{sec:spin 4-vector} and \secref{sec:T-BMT equations} by making the replacement
\begin{subequations}\label{replacement}
\begin{eqnarray}
U^\alpha,\ \mathbf{v},\ \gamma &\longrightarrow&
U^\alpha_{\boldsymbol{\pi}},\ \mathbf{v}_{\boldsymbol{\pi}},\ \gamma_{\boldsymbol{\pi}},\\
d\tau\equiv\gamma^{-1}dt &\longrightarrow&
d\tau_{\boldsymbol{\pi}}:=\gamma_{\boldsymbol{\pi}}^{-1}dt
\end{eqnarray}
\end{subequations}
on every occasion and obtain \eqnref{TMBT new} in the end.
This seems to suggest that the Pauli matrices $\sigma_i$ used in the Dirac-Pauli equation represent the intrinsic spin vector (via $\mathbf{s}=\hbar\boldsymbol{\sigma}/2$) for the ``rest frame'' comoving with $\mathbf{v}_{\boldsymbol{\pi}}$, instead of $\mathbf{v}$, and accordingly the spin 4-vector satisfies the constraint
\begin{equation}\label{constraint 1 modified}
U^\alpha_{\boldsymbol{\pi}}S_\alpha=0,
\end{equation}
instead of \eqnref{constraint 1}, although the difference is negligible if $H_\mathrm{spin}\ll mc^2$.

Finally, with \eqnref{H total} modified to \eqnref{H total modified}, by repeating the calculations towards \eqnref{Lorentz eq}, Hamilton's equation $d\mathbf{p}/dt=-\boldsymbol{\nabla} H$ now gives
\begin{eqnarray}\label{Lorentz eq modified}
&&\frac{d\boldsymbol{\pi}}{d\tau_{\boldsymbol{\pi}}}
\equiv\gamma_{\boldsymbol{\pi}}\frac{d\left(\mathbf{p}-\frac{e}{c}\mathbf{A}\right)}{dt}\nonumber\\
&=&\frac{e}{c}\left[(\gamma_{\boldsymbol{\pi}}c)\mathbf{E}
+(\gamma_{\boldsymbol{\pi}}\mathbf{v}_{\boldsymbol{\pi}})\times\mathbf{B}\right]\nonumber\\
&&\mbox{}
-\frac{e}{c}\gamma_{\boldsymbol{\pi}}
\left(\left(\mathbf{v}-\mathbf{v}_{\boldsymbol{\pi}}\right)\cdot\boldsymbol{\nabla}\right)\mathbf{A}
-\gamma_{\boldsymbol{\pi}}\boldsymbol{\nabla} H_\mathrm{spin}\nonumber\\
&&\mbox{}-\gamma_{\boldsymbol{\pi}}\boldsymbol{\nabla}\left(\mathcal{O}(F_{\mu\nu}^2,\hbar^2)\right),
\end{eqnarray}
which is lifted to the tensorial form
\begin{equation}
m\frac{dU_{\boldsymbol{\pi}}^\alpha}{d\tau_{\boldsymbol{\pi}}}
=\frac{e}{c}F^{\alpha\beta}U_{\boldsymbol{\pi}\beta}
+f^\alpha
\end{equation}
as \eqnref{covariant Lorentz eq} prescribed with \eqnref{replacement} as desired. However, it should be caveated that the second and third lines of \eqnref{Lorentz eq modified} together do not give rise to spatial components of a 4-vector $f^\alpha$, unless $\mathcal{O}(F_{\mu\nu}^2,\hbar^2)$ are appropriately supplemented.
This tells us that the consistency with Lorentz covariance relies on a more fundamental theory and, rigourously speaking, the classical theory described by \eqnref{H total modified} respects Lorentz invariance only within a high degree of accuracy if ``$\mathcal{O}(F_{\mu\nu}^2,\hbar^2)$'' is omitted.
Indeed, as we will see in \secref{sec:Dirac-Pauli spinor}, the Dirac-Pauli equation, regarded as the underlying fundamental theory for charged spinors, does give rise to (infinitely many) nonlinear electromagnetic corrections of $\mathcal{O}(F_{\mu\nu}^2)$ as well as the Darwin term of $\mathcal{O}(\hbar^2)$ in addition to the sum of orbital and spin Hamiltonians.
Furthermore, as in \eqnref{Dirac eq 2} with \eqnref{covariant dev}, the Dirac equation leads to the relativistic dispersion relation
\begin{equation}
W^2=\boldsymbol{\pi}^2c^2+m^2c^4,
\end{equation}
where $W=E-e\phi$ is the kinematic energy while $E$ is the canonical energy, and $\boldsymbol{\pi}$ is the kinematic momentum defined in \eqnref{def pi}. Consequently, the Dirac equation dictates that the 4-vector of kinematic momentum given by
\begin{eqnarray}
\pi^\alpha&:=&p^\alpha-\frac{e}{c}A^\alpha\nonumber\\
&=&\left(\frac{E-e\,\phi}{c},\,\mathbf{p}-\frac{e}{c}\,\mathbf{A}\right)
\equiv\left(\frac{W}{c},\boldsymbol{\pi}\right)
\end{eqnarray}
yields
\begin{equation}\label{pi square}
\pi^\alpha\pi_\alpha=m^2c^2,
\end{equation}
which justifies $U_{\boldsymbol{\pi}}$ defined in \eqnref{U pi} to be a 4-vector with
\begin{equation}
\pi^\alpha=mU_{\boldsymbol{\pi}}^\alpha.
\end{equation}
If, however, the fundamental theory is governed by the Dirac-Pauli equation, instead of the Dirac equation, the right-hand side of \eqnref{pi square} is no longer constant but will receive corrections for interactions between spin and $F_{\mu\nu}$, since \eqnref{Dirac-Pauli eq} does not lead to \eqnref{Dirac eq 2} but will introduce extra corrections. The corrections do not change \eqnref{T-BMT new F} if we only consider $\mathbf{E}$ and $\mathbf{B}$ to the linear order in the weak-field limit.

\section{Dirac-Pauli spinor}\label{sec:Dirac-Pauli spinor}
Now we move to the second part for the relativistic quantum dynamics of the Dirac-Pauli spinor. After briefly reviewing the Dirac and Dirac-Pauli equations, we apply Eriksen's method to obtain the exact FW transformed Hamiltonian for two special cases. The two special cases considered together suggest that, in the weak-field limit, the FW transformed Dirac-Pauli Hamiltonian agrees with the classical Hamiltonian \eqnref{H total modified}--\eqnref{T-BMT new F} plus the Darwin term, and the canonical variables in \eqnref{H total modified} are promoted to corresponding quantum operators via a specific way of Weyl ordering. The Darwin term is shown to have no classical (non-quantum) correspondence.

\subsection{Dirac-Pauli equation}
The relativistic quantum theory of a spin-$1/2$ particle subject to external electromagnetic fields is described by the Dirac equation \cite{Dirac1928,Dirac1982}
\begin{equation}\label{Dirac eq}
\mgamma^\mu D_\mu|\psi\rangle+i\frac{mc}{\hbar}|\psi\rangle=0,
\end{equation}
where the Dirac bispinor $|\psi\rangle=(\chi,\varphi)^T$ is composed of two 2-component Weyl spinors $\chi$ and $\varphi$, the covariant derivative $D_\mu$ is given by
\begin{eqnarray}\label{covariant dev}
D_\mu&:=&\partial_\mu +\frac{ie}{\hbar c}A_\mu \equiv -\frac{i}{\hbar}\pi_\mu :=-\frac{i}{\hbar}\left(p_\mu-\frac{e}{c}A_\mu\right)\nonumber\\
&=&\left(\frac{1}{c}\frac{\partial}{\partial t}+\frac{ie}{\hbar c}\phi,\,\boldsymbol{\nabla}-\frac{ie}{\hbar c}\mathbf{A}\right)\nonumber\\
&\equiv& -\frac{i}{\hbar}\left(\frac{E-e\,\phi}{c}, \,-\left(\mathbf{p}-\frac{e}{c}\mathbf{A}\right)\right)
\end{eqnarray}
with $p^\mu=(E/c,\mathbf{p})$ being the 4-vector of canonical energy and momentum and $\pi^\mu=(W/c,\boldsymbol{\pi})$ being the 4-vector of kinematic energy and momentum, and $\mgamma^\mu$ are $4\times4$ matrices\footnote{A tilde is attached to denote a $4\times4$ Dirac matrix.} which satisfy
\begin{equation}
\mgamma^\mu\mgamma^\nu+\mgamma^\nu\mgamma^\mu = 2g^{\mu\nu}.
\end{equation}
It is easy to show that \eqnref{Dirac eq} leads to the wave function
\begin{equation}\label{Dirac eq 2}
\hbar^2D_\mu D^\mu \psi +m^2c^2\psi=0.
\end{equation}
The Dirac equation gives rise to the magnetic moment with $\gamma_m=e/(mc)$ (i.e., $g=2$). To incorporate any anomalous magnetic moment $\mu'$ (i.e., $\gamma_m= {e}/{(mc)}+2{\mu'}/{\hbar}$),\footnote{Also see \footref{foot:gyromagnetic ratio}.} one can modify the Dirac equation to the Dirac-Pauli equation with augmentation of explicit dependence on field strength \cite{Silenko2008,Pauli1941}:
\begin{equation}\label{Dirac-Pauli eq}
\mgamma^\mu D_\mu|\psi\rangle+i\frac{mc}{\hbar}|\psi\rangle
+\frac{i\mu'}{2c}\mgamma^\mu\mgamma^\nu F_{\mu\nu}|\psi\rangle=0.
\end{equation}

The Pauli-Dirac equation can be cast in Hamiltonian formalism as
\begin{equation}\label{Dirac-Pauli eq w H}
i\hbar\frac{\partial}{\partial t}|\psi\rangle = H|\psi\rangle
\end{equation}
with the Dirac-Pauli Hamiltonian:
\begin{eqnarray}\label{H Dirac-Pauli}
H &=& mc^2\mbeta+c\,\mbalpha\cdot\left(\mathbf{p}-\frac{e}{c}\mathbf{A}\right)+e\,\phi\nonumber\\
&&\mbox{} +\mu'\left(-\mbeta\mbsigma\cdot\mathbf{B}+i\mbeta\mbalpha\cdot\mathbf{E}\right),
\end{eqnarray}
where the $4\times4$ matrices are given explicitly by
\begin{equation}
\mbeta=\left(
         \begin{array}{cc}
           \openone & 0 \\
           0 & -\openone \\
         \end{array}
       \right),\
\mbalpha=\left(
           \begin{array}{cc}
             0 & \boldsymbol{\sigma} \\
             \boldsymbol{\sigma} & 0 \\
           \end{array}
         \right),\
\mbsigma=\left(
           \begin{array}{cc}
             \boldsymbol{\sigma} & 0 \\
             0 & \boldsymbol{\sigma} \\
           \end{array}
         \right),
\end{equation}
and $\boldsymbol{\sigma}=(\sigma_x,\sigma_y,\sigma_z)$ are the $2\times2$ Pauli matrices.
Accordingly, the $\mgamma$ matrices are given by
\begin{equation}
\mgamma^0=\mbeta,
\quad
\mgamma^i=\mbeta\malpha^i=\left(
           \begin{array}{cc}
             0 & \sigma_i \\
             -\sigma_i & 0 \\
           \end{array}
         \right).
\end{equation}

\subsection{Foldy-Wouthuysen transformation}
The Dirac-Pauli Hamiltonian \eqnref{H Dirac-Pauli} (or, more generally, with other corrections) can be schematically put in the form
\begin{equation}
H=\mbeta mc^2+\mathcal{O}+\mathcal{E},
\end{equation}
where $\mathcal{E}$ is the ``even'' part which commutes with $\mbeta$, i.e., $\mbeta\mathcal{E}\mbeta=\mathcal{E}$, while $\mathcal{O}$ is the ``odd'' part which anticommutes with $\mbeta$, i.e., $\mbeta\mathcal{O}\mbeta=-\mathcal{O}$. Because of the presence of the odd part, the Hamiltonian in the Dirac bispinor representation is not block diagonalized, and thus the particle and antiparticle components are entangled in each of the Weyl spinors $\chi$ and $\varphi$. The question that naturally arises is whether we can find a representation in which the particle and antiparticle are separated, or equivalently, the Hamiltonian is block-diagonalized. Foldy and Wouthuysen have shown that such a representation is possible \cite{Foldy1950,Strange2008}. The Foldy-Wouthuysen (FW) transformation is a unitary and nonexplicitly time-dependent transformation on the Dirac bispinor
\begin{subequations}
\begin{eqnarray}
|\psi\rangle&\rightarrow&|\psi'\rangle=U_\mathrm{FW}|\psi\rangle,\\
\label{H'}
H'&=&U_\mathrm{FW}HU^\dagger_\mathrm{FW},
\end{eqnarray}
\end{subequations}
which leaves \eqnref{Dirac-Pauli eq w H} in the form
\begin{equation}\label{H' psi'}
i\hbar\frac{\partial}{\partial t}|\psi'\rangle = H'|\psi'\rangle
\end{equation}
and block-diagonalizes the Hamiltonian, i.e.
\begin{equation}
[\mbeta,H']=0.
\end{equation}

As the FW transformation separates the particle and antiparticle components, the two diagonal blocks of $H'$ are adequate to describe the relativistic quantum dynamics of the spin-1/2 particle and antiparticle respectively in the low-energy regime in which the field-theory corrections due to particle-antiparticle pair creation/annihilation are negligible. It is then natural to conjecture that each diagonal block of $H'$ in the low-energy/weak-field limit agrees with its classical counterpart, i.e., the classical orbital Hamiltonian plus the spin Hamiltonian, as given in \eqnref{H total modified}--\eqnref{T-BMT new F}.
For the case of static and homogeneous electromagnetic fields, in the weak-field limit, the conjecture is shown to be true up to the order of $1/m^{14}$ \cite{Chen2010,Chang:thesis,paper-to-appear}.

When the applied electromagnetic field is static but inhomogeneous, the issue of operator ordering arises when we promote canonical variables in \eqnref{H total modified} to quantum operators. As will be shown in \secref{sec:Eriksen's method}, in the low-energy/weak-field limit in which nonlinear electromagnetic effects are neglected, it is strongly suggested that, in accordance with \eqnref{H total modified}--\eqnref{T-BMT new F} with $\mathbf{s}=\hbar\boldsymbol{\sigma}/2$, the exact FW transform of the Dirac-Pauli Hamiltonian is given by
\begin{eqnarray}\label{H total FW}
&&H'(\mathbf{x},\mathbf{p})\nonumber\\
&=&\mbeta\sqrt{c^2\boldsymbol{\pi}^2+m^2c^4}\,+e\,\phi(\mathbf{x})\nonumber\\
&&
\mbox{}
-\mbeta\,\frac{\hbar}{2}\,\mbsigma\cdot
\Bigg[
\left(\gamma_m-\frac{e}{mc}
+\frac{e}{mc}\frac{1}{\gamma_{\boldsymbol{\pi}}}\right)\mathbf{B}(\mathbf{x})
\nonumber\\
&&\qquad\qquad\quad
-\left(\gamma_m-\frac{e}{mc}\right)
\frac{\overline{\left(\boldsymbol{\pi}\cdot\mathbf{B}(\mathbf{x})\right)\boldsymbol{\pi}}}
{\gamma_{\boldsymbol{\pi}}(\gamma_{\boldsymbol{\pi}}+1)m^2c^2}
\nonumber\\
&&\qquad\qquad\quad
-\mbeta\left(\gamma_m-\frac{e}{mc}
\frac{\gamma_{\boldsymbol{\pi}}}{\gamma_{\boldsymbol{\pi}}+1}\right)
\frac{\overline{\boldsymbol{\pi}\times\mathbf{E}(\mathbf{x})}}{\gamma_{\boldsymbol{\pi}}mc}
\Bigg]_\mathrm{Weyl}\nonumber\\
&&\mbox{}
+\frac{\hbar^2}{4mc}\left(\frac{3\,e}{2mc}-\gamma_m\right)
\left(\frac{\boldsymbol{\nabla}\cdot\mathbf{E}(\mathbf{x})}{\gamma_{\boldsymbol{\pi}}}
\right)_\mathrm{Weyl},
\end{eqnarray}
where
\begin{eqnarray}\label{pi dot F pi}
&&\overline{(\boldsymbol{\pi}\cdot\mathbf{F})\boldsymbol{\pi}}\\
&:=&\frac{1}{4}\left(\boldsymbol{\pi}\cdot\mathbf{F}
+\mathbf{F}\cdot\boldsymbol{\pi}\right)\boldsymbol{\pi}
+\frac{1}{4}\boldsymbol{\pi}\left(\boldsymbol{\pi}\cdot\mathbf{F}
+\mathbf{F}\cdot\boldsymbol{\pi}\right)\nonumber
\end{eqnarray}
is the operator symmetrization of $(\boldsymbol{\pi}\cdot\mathbf{F})\boldsymbol{\pi}$ (with $\mathbf{F}$ being $\mathbf{E}$ or $\mathbf{B}$),
\begin{equation}\label{pi times F}
\overline{\boldsymbol{\pi}\times\mathbf{F}}\equiv-\overline{\mathbf{F}\times\boldsymbol{\pi}}
:=\frac{1}{2}\left(\boldsymbol{\pi}\times\mathbf{F}-\mathbf{F}\times\boldsymbol{\pi}\right)
\end{equation}
is the operator symmetrization of $\boldsymbol{\pi}\times\mathbf{F}$, and the subscript ``Weyl'' denotes the specific Weyl ordering performed over $\mathbf{B}$, $\overline{(\boldsymbol{\pi}\cdot\mathbf{B})\boldsymbol{\pi}}$, $\overline{\boldsymbol{\pi}\times\mathbf{E}}$, and $\boldsymbol{\nabla}\cdot\mathbf{E}$ with powers of $\boldsymbol{\pi}^2$ (which arises from the Taylor series of functions of $\gamma_{\boldsymbol{\pi}}$):
\begin{equation}\label{Weyl ordering}
\left(\mathbf{X}\,\boldsymbol{\pi}^{2n}\right)_\mathrm{Weyl} :=\frac{1}{n+1}\sum_{l=0}^n\boldsymbol{\pi}^{2l}\,\mathbf{X}\,\boldsymbol{\pi}^{2n-2l}
\end{equation}
for $\mathbf{X}$ being $\mathbf{B}$, $\overline{(\boldsymbol{\pi}\cdot\mathbf{B})\boldsymbol{\pi}}$, $\overline{\boldsymbol{\pi}\times\mathbf{E}}$, or $\boldsymbol{\nabla}\cdot\mathbf{E}$. Note that, in \eqnref{H total FW}, $\sqrt{c^2\boldsymbol{\pi}^2+m^2c^4}=\gamma_{\boldsymbol{\pi}}mc^2$ and functions of the operator $\gamma_{\boldsymbol{\pi}}$, whose classical counterpart is defined in \eqnref{gamma pi}, are understood via the Taylor series for the function of an operator $\Omega$ as
\begin{equation}\label{Taylor series}
f\left(1+\frac{\Omega}{m^2c^2}\right)
=\sum_{n=0}^{\infty}\frac{f^{(n)}(1)}{n!}\left(\frac{\Omega}{m^2c^2}\right)^n,
\end{equation}
which produces convergent results provided the spectrum of $\Omega$ satisfies $|\Omega|<m^2c^2$ (namely, the relevant energy given by $\sqrt{|\Omega c^2|}$ is less than the energy gap $2mc^2$ for particle-antiparticle pair creation).

Also note that, when promoted from the classical Hamiltonian \eqnref{H total modified}--\eqnref{T-BMT new F} to the FW transformed Dirac-Pauli Hamiltonian \eqnref{H total FW}, in additional to the operator ordering, the matrix $\mbeta$ is also supposed to appear in front of various terms such that $\mbeta H'$ is invariant under charge conjugation (C), parity (P), and time reversal (T). That is, $\mbeta H'$ with \eqnref{H total FW} is invariant under sign flips specified in each column in \tabref{tab:symmetries}.\footnote{Note that occurrences of $\mbeta$ in \eqnref{H total FW} are in accord with those in \eqnref{H' standard}, which is given by (7.111) of \cite{Strange2008}. See $\S$6 of \cite{Strange2008} for more on the CPT symmetries.}

\begin{table}
\begin{tabular}{c|cccc}
                & \;C\; & \;P\; & \;T\; & \;CPT\; \\
  \hline
  $e,\gamma_m$  & $-$ & $+$ & $+$ & $-$ \\
  $\mathbf{x},\boldsymbol{\nabla}$
                & $+$ & $-$ & $+$ & $-$ \\
  $\mathbf{p},\boldsymbol{\pi}$
                & $-$ & $-$ & $-$ & $-$ \\
  $\phi$        & $+$ & $+$ & $+$ & $+$ \\
  $\mathbf{A}$  & $+$ & $-$ & $-$ & $+$ \\
  $\mathbf{E}=-\boldsymbol{\nabla}\phi+c^{-1}\partial_t\mathbf{A}$
                & $+$ & $-$ & $+$ & $-$ \\
  $\mathbf{B}=\boldsymbol{\nabla}\times\mathbf{A}$
                & $+$ & $+$ & $-$ & $-$ \\
  $\mbsigma=-i\mbalpha\times\mbalpha/2$
                & $-$ & $+$ & $-$ & $+$ \\
  $\mbeta$      & $-$ & $+$ & $+$ & $-$
\end{tabular}
\caption{\label{tab:symmetries}Symmetries of various operators and physical quantities under charge conjugation (C), parity (P) and time reversal (T).}
\end{table}

Finally, the last term in \eqnref{H total FW} is included and known as the \emph{Darwin term}, which comes out from \emph{Zitterbewegung} and has no direct classical correspondence. More about it will be remarked in \secref{sec:Darwin term}.

\subsection{Eriksen's method}\label{sec:Eriksen's method}
It is tremendously difficult to prove \eqnref{H total FW} to a high order in an order-by-order scenario when the electromagnetic field is no longer homogeneous, as complications arise from operator ordering. Fortunately, by the method proposed by Eriksen \cite{Eriksen1958}, the exact FW transformation can be found if the Dirac-Pauli Hamiltonian is not explicitly time dependent and involves only the odd part, i.e.,
\begin{equation}\label{H odd}
H=\mbeta mc^2+\mathcal{O}.
\end{equation}
The exact FW transform of $H$ above is given by
\begin{equation}\label{H' exact}
H'=U_\mathrm{FW}HU^\dagger_\mathrm{FW}
=\mbeta\left[m^2c^4+\mathcal{O}^2\right]^{1/2},
\end{equation}
where again the function of the operator $\mathcal{O}$ is understood via \eqnref{Taylor series}.
Although the Dirac-Pauli Hamiltonian \eqnref{H Dirac-Pauli} does not fit into the generic form of \eqnref{H odd}, we can investigate two special cases of the form which are complementary to each other and, when combined together, suggest the complete expression of \eqnref{H total FW}. In the following, as we are interested only in the low-energy limit, we assume the applied electromagnetic fields to be weak enough and ignore terms nonlinear in electromagnetic fields.

\subsubsection{Special case I}
As the first special case, let us consider a Dirac spinor ($\mu'=0$) with charge $e$ subject to a magnetostatic field ($\partial_t\mathbf{B}=0$ and $\mathbf{E}=0$). The Dirac-Pauli Hamiltonian \eqnref{H Dirac-Pauli} then reads as
\begin{equation}
H=\mbeta mc^2+\mbalpha\cdot(c\,\mathbf{p}-e\mathbf{A})=\mbeta mc^2+\mathcal{O}.
\end{equation}
Applying the identity
\begin{equation}\label{alpha identity}
\mbalpha\cdot\mathbf{A}\, \mbalpha\cdot\mathbf{B} =\mathbf{A}\cdot\mathbf{B}+i\mbsigma\cdot\mathbf{A}\times\mathbf{B},
\end{equation}
\eqnref{H' exact} then yields\footnote{Note that $(\boldsymbol{\nabla}\times\mathbf{A}+\mathbf{A}\times\boldsymbol{\nabla})\psi =(\boldsymbol{\nabla}\times\mathbf{A})\psi=\mathbf{B}\,\psi$.}
\begin{equation}\label{exact H' I}
H'=\mbeta\left[m^2c^4+c^2\boldsymbol{\pi}^2-e\hbar c\,\mbsigma\cdot\mathbf{B}\right]^{1/2}.
\end{equation}

Define the operator
\begin{equation}
\Omega=\boldsymbol{\pi}^2-\frac{e\hbar}{c}\,\mbsigma\cdot\mathbf{B}.
\end{equation}
Neglecting quadratic and higher-order terms in $\mathbf{B}$, it is easy to show by induction that
\begin{equation}
\Omega^n=\boldsymbol{\pi}^{2n}
-n \left(\frac{e\hbar}{c}\,\mbsigma\cdot\mathbf{B}\,\boldsymbol{\pi}^{2n-2}\right)_\mathrm{Weyl},
\end{equation}
where the Weyl ordering is defined in \eqnref{Weyl ordering}.
Consequently, we obtain\footnote{The identities $(1+x)^{1/2}=\sum_{n=0}^\infty {\frac{1}{2} \choose n} x^n$ and $\frac{1}{2}(1+x)^{-1/2}=\sum_{n=0}^\infty {\frac{1}{2} \choose n}n\, x^{n-1}$ are used.}
\begin{eqnarray}\label{H' case I}
H' &=& \mbeta mc^2\left[1+\frac{\Omega}{m^2c^2}\right]^{1/2}
 \equiv \mbeta mc^2\sum_{n=0}^\infty {\frac{1}{2} \choose n}
 \left(\frac{\Omega}{m^2c^2}\right)^n\nonumber\\
 &=& \mbeta \left[
 \sqrt{m^2c^4+c^2\boldsymbol{\pi}^2} -\frac{e\hbar}{2mc}\, \frac{\mbsigma\cdot\mathbf{B}}{\sqrt{1+\left(\frac{\boldsymbol{\pi}}{mc}\right)^2}}
 \right]_\mathrm{Weyl}\nonumber\\
 &=& \mbeta \left[
 \sqrt{m^2c^4+c^2\boldsymbol{\pi}^2}
 -\frac{e\hbar}{2mc}\frac{\mbsigma\cdot\mathbf{B}}{\gamma_{\boldsymbol{\pi}}}
 \right]_\mathrm{Weyl},
\end{eqnarray}
which is in full agreement with \eqnref{H total FW} by identifying $\mathbf{E}=0$ and
\begin{equation}\label{gamma m g=2}
\gamma_m = \frac{e}{mc}.
\end{equation}

Even though the case we considered above is restricted to $\mathbf{E}=0$ (and $\phi=0$), any effects involved with $\phi$ and $\mathbf{E}$ (except the Darwin term) will arise if a Lorentz boost is performed. Because the Dirac-Pauli equation \eqnref{Dirac-Pauli eq} respects the Lorentz invariance explicitly, and meanwhile the corresponding classical Hamiltonian given by \eqnref{H total modified}--\eqnref{T-BMT new F} is obtained by Lorentz covariance (as derived in \secref{sec:classical relativistic spinor}), it is therefore anticipated that nonzero $\phi$ and $\mathbf{E}$ produced by the boost will give rise to the electric energy as the term of $e\phi$ in \eqnref{H total FW} and the spin-orbit interaction energy as the third term inside the square brackets in \eqnref{H total FW} (with \eqnref{gamma m g=2} imposed).

On the other hand, the second term inside the square brackets in \eqnref{H total FW}, which is responsible for the precession of longitudinal polarization, vanishes identically because the Dirac spinor gives \eqnref{gamma m g=2}, i.e., $g=2$. In order to see this term in quantum theory, we consider the Dirac-Pauli spinor in the second case to include an anomalous magnetic moment.

\subsubsection{Special case II}
As the second special case, let us consider a Dirac-Pauli spinor with zero charge ($e=0$) but nonzero magnetic moment $\mu'$ subject to an electrostatic field ($\partial_t\mathbf{E}=0$ and $\mathbf{B}=0$). The Dirac-Pauli Hamiltonian \eqnref{H Dirac-Pauli} now reads as
\begin{equation}
H=\mbeta mc^2+\mbalpha\cdot(c\,\mathbf{p})+i\mu'\mbeta\,\mbalpha\cdot\mathbf{E}=\mbeta mc^2+\mathcal{O}.
\end{equation}
Applying \eqnref{alpha identity}, \eqnref{H' exact} then yields\footnote{Note that $(\mathbf{p}\cdot\mathbf{E}-\mathbf{E}\cdot\mathbf{p})\psi =\frac{\hbar}{i}(\boldsymbol{\nabla}\cdot\mathbf{E})\psi$.}
\begin{eqnarray}\label{exact H' II}
H'&=&\mbeta\Big[m^2c^4+c^2\mathbf{p}^2-\mu'\hbar c \,\mbeta\,\boldsymbol{\nabla}\cdot\mathbf{E}\nonumber\\
&&\qquad+2\mu' c\,\beta\,\mbsigma\cdot\overline{\mathbf{p}\times\mathbf{E}} +\mu'^2\mathbf{E}^2
\Big]^{1/2},
\end{eqnarray}
where $\overline{\mathbf{p}\times\mathbf{B}}$ is defined in \eqnref{pi times F}.

Define the operator
\begin{equation}
\Omega=\boldsymbol{\pi}^2-\frac{\mu'\hbar}{c}\,\mbeta\,\boldsymbol{\nabla}\cdot\mathbf{E}
+\frac{2\mu'}{c}\,\mbeta\,\mbsigma\cdot\overline{\boldsymbol{\pi}\times\mathbf{E}},
\end{equation}
where $\boldsymbol{\pi}=\mathbf{p}$.
Neglecting quadratic and higher-order terms in $\mathbf{E}$, it is easy to show by induction that
\begin{eqnarray}
\Omega^n&=&\boldsymbol{\pi}^{2n}
-n
\left[\frac{\mu'\hbar}{c}\,\mbeta\,(\boldsymbol{\nabla}\cdot\mathbf{E})\,\boldsymbol{\pi}^{2n-2}
\right.\\
&&\qquad\qquad\quad
\left.
-\frac{2\mu'}{c}\,\mbeta\,\mbsigma\cdot\overline{\boldsymbol{\pi}\times\mathbf{E}}\,\boldsymbol{\pi}^{2n-2}
\right]_\mathrm{Weyl}.\nonumber
\end{eqnarray}
Consequently, we obtain
\begin{eqnarray}\label{H' case II}
H' &=& \mbeta mc^2\left[1+\frac{\Omega}{m^2c^2}\right]^{1/2}
\equiv \mbeta mc^2\sum_{n=0}^\infty {\frac{1}{2} \choose n} \left(\frac{\Omega}{m^2c^2}\right)^n\nonumber\\
&=& \mbeta \left[
\sqrt{m^2c^4+c^2\boldsymbol{\pi}^2} -\frac{\mu'\hbar}{2mc}\,\mbeta\,(\boldsymbol{\nabla}\cdot\mathbf{E})
\,\frac{1}{\gamma_{\boldsymbol{\pi}}}\right.\nonumber\\
&&\qquad
\left.
+\frac{\mu'}{mc}\,\mbeta\,\mbsigma\cdot\overline{\boldsymbol{\pi}\times\mathbf{E}}
\,\frac{1}{\gamma_{\boldsymbol{\pi}}}
\right]_\mathrm{Weyl},
\end{eqnarray}
which is in full agreement with \eqnref{H total FW} including the Darwin term by identifying $\mathbf{B}=0$, $e=0$, and
\begin{equation}\label{gamma m mu}
\gamma_m = \frac{2\mu'}{\hbar}.
\end{equation}

Again, even though the case we considered above is restricted to $\mathbf{B}=0$ (and $\mathbf{A}=0$), any effects involved with $\mathbf{B}$ and $\mathbf{A}$ will arise if a Lorentz boost is performed. By Lorentz covariance, it is anticipated that nonzero $\mathbf{B}$ produced by the boost will give rise to the first and second terms inside the square brackets in \eqnref{H total FW} (with $e=0$ and \eqnref{gamma m mu} imposed). In this case, the second term for the precession of longitudinal polarization is no longer identically zero.

The two special cases given above in conjunction suggest the complete form of \eqnref{H total FW} except for the Darwin term (which will be discussed in \secref{sec:Darwin term}), if Lorentz covariance under Lorentz boosts is taken into consideration. The only thing not completely certain is the detailed operator ordering for the second term inside the square brackets in \eqnref{H total FW}, but the prescription given by \eqnref{pi dot F pi} is the most natural one as will be discussed in \secref{sec:ordering}.

Furthermore, under a Lorentz boost, inhomogeneity of electromagnetic fields will give rise to nonstaticity. Accordingly, the Darwin term involving $\boldsymbol{\nabla}\cdot\mathbf{E}$ is expected to have the counterparts involving $\partial_t\mathbf{E}$ and $\boldsymbol{\nabla}\times\mathbf{B}$, but discussions in \secref{sec:Darwin term} suggest this is not the case.

Meanwhile, it should be noted that the exact FW transformed Hamiltonian, as in \eqnref{exact H' I} and \eqnref{exact H' II}, contains infinitely many nonlinear electromagnetic corrections of $\mathcal{O}(F_{\mu\nu}^2)$, which are neglected in the weak-field limit. The terms of $\mathcal{O}(F_{\mu\nu}^2)$ and of $\mathcal{O}(\hbar^2)$ (including the Darwin term) give rise to $\mathcal{O}(F_{\mu\nu}^2,\hbar^2)$ in \eqnref{H total modified}. As commented earlier in \secref{sec:Hamiltonian formalism}, occurrence of $\mathcal{O}(F_{\mu\nu}^2,\hbar^2)$ terms is an inevitable consequence of Lorentz invariance.

\subsection{Remarks on the Darwin term}\label{sec:Darwin term}
The standard procedure of performing a series of successive FW transformations on the Dirac Hamiltonian yields (as shown in (7.111) of \cite{Strange2008})
\begin{eqnarray}\label{H' standard}
H'&=&\mbeta mc^2 +\mbeta\frac{\boldsymbol{\pi}^2}{2m} -\mbeta\frac{\boldsymbol{\pi}^4}{8m^3c^2} +e\,\phi\\
&&\mbox{}-\frac{e\hbar}{2m}\mbeta\,\mbsigma\cdot\mathbf{B}
+\frac{\hbar e}{4m^2c^2}\,\mbsigma\cdot\boldsymbol{\pi}\times\mathbf{E}
+\frac{\hbar^2e}{8m^2c^2}\boldsymbol{\nabla}\cdot\mathbf{E}
\nonumber
\end{eqnarray}
to the order of $1/m^4$. Note that the Darwin term is given by $\frac{\hbar^2e}{8m^2c^2}\boldsymbol{\nabla}\cdot\mathbf{E}$, which does not appear in \eqnref{H' case I} for the first special case since $\mathbf{E}$ is set to be zero. On the other hand, for the Dirac-Pauli Hamiltonian, the second special case does gives rise to the Darwin term in \eqnref{H' case II}, which reduces to $-\frac{\mu'\hbar}{2mc}(\boldsymbol{\nabla}\cdot\mathbf{E})$ if corrections of the inverse Lorentz $\gamma$ factor are neglected. Put together, the Darwin term for the Dirac-Pauli spinor (with arbitrary charge $e$ and gyromagnetic ratio $\gamma_m$) is conjectured to take the form of the last term in \eqnref{H total FW}.

Equations \eqnref{gamma m g=2} and \eqnref{gamma m mu} imply that the magnetic moment of a Dirac-Pauli spinor is given by
\begin{equation}\label{magnetic moment}
\mu:=\gamma_m\frac{\hbar}{2}=\frac{e\hbar}{2mc}+\mu',
\end{equation}
where the first part is the Dirac magnetic moment (with $g$-factor $g=2$) and $\mu'$ is the anomalous magnetic moment.
When $e$ and $\mu'$ are of the same sign, the magnitude of the magnetic moment is increased by the presence of $\mu'$. However, it should be noted that the Darwin term, by contrast, is decreased by the presence of $\mu'$, as can be seen by substituting \eqnref{magnetic moment} into the last term in \eqnref{H total FW}.

The Darwin term comes from the fact that the Dirac or Dirac-Pauli spinor cannot be regarded as a point particle but is oscillating extremely rapidly and thus is spread out over a volume of the order of the cube of the Compton wavelength $(\hbar/mc)^3$. The rapid oscillatory motion is called \emph{Zitterbewegung}. (See \cite{Strange2008} for a review.)

Although the Darwin term has no classical (non-quantum) analog, we wonder whether we can phenomenologically include its effects at the level of classical equations of motion such that Lorentz covariance is upheld.
In accordance with the Darwin term in \eqnref{H total FW}, the classical equation of the orbital motion \eqnref{covariant Lorentz eq} is given by
\begin{equation}
\frac{d\pi^\alpha}{d\tau}=\frac{e}{c}F^{\alpha\beta}U_\beta -\partial^\alpha \mathscr{H}_D+\cdots,
\end{equation}
where
\begin{equation}
\mathscr{H}_D=A_D\partial_\nu F^{\mu\nu}U_\mu
\end{equation}
is the relativistic correspondence to $A_D\boldsymbol{\nabla}\cdot\mathbf{E}$ with a constant $A_D$. This leads to
\begin{equation}
\frac{d\boldsymbol{\pi}}{dt}
=e\left(\mathbf{E}+\frac{\mathbf{v}}{c}\times\mathbf{B}\right) -\boldsymbol{\nabla}H_D+\cdots
\end{equation}
with
\begin{equation}\label{H D}
H_D=cA_D\left(\boldsymbol{\nabla}\cdot\mathbf{E}
+\frac{\mathbf{v}}{c^2}\cdot\partial_t\mathbf{E}
-\frac{\mathbf{v}}{c}\cdot\left(\boldsymbol{\nabla}\times\mathbf{B}\right)
\right).
\end{equation}
The Hamiltonian \eqnref{H D} however is not the classical correspondence to the Darwin term. If it were, in accord with the third term of \eqnref{H D}, we would have obtained the additional Darwin term involving with $\boldsymbol{\nabla}\times\mathbf{B}$ in \eqnref{H' case I} for the first special case, and in accord with the first term of \eqnref{H D}, we would not have had the $1/\gamma_{\boldsymbol{\pi}}$ factor for the Darwin term in \eqnref{H' case II} for the second special case.
Therefore, we realize that the Darwin term is essentially a relativistic quantum effect and has no non-quantum correspondence.

In the quantum theory, we are uncertain whether the Darwin term has a counterpart involving $\partial_t\mathbf{E}$ and/or $\partial_t\mathbf{B}$, because if the unitary transformation $U_\mathrm{FW}$ is explicitly time dependent, instead of \eqnref{H'}, the diagonalized Hamiltonian is given by
\begin{equation}
H'=U_\mathrm{FW}HU^\dagger_\mathrm{FW}-i\hbar\, U_\mathrm{FW}\frac{\partial}{\partial t} U^\dagger_\mathrm{FW},
\end{equation}
which is beyond the scope of the standard FW scenario, including Eriksen's method.

\subsection{Remarks on the operator ordering}\label{sec:ordering}
As commented previously, since we consider only two special cases, the precise operator ordering for the second term inside the square brackets in \eqnref{H total FW} cannot be assured completely. Nevertheless, heuristic considerations suggest that the correct operator ordering should take the form prescribed by \eqnref{pi dot F pi}.

Under a Lorentz transformation from a reference $K$ to another reference $K'$ (not necessarily the particle's rest frame) moving with boost velocity $\mathbf{v}$ relative to $K$, the electromagnetic field transforms as a rank-2 antisymmetric tensor:
\begin{subequations}\label{Lorentz trans of E B}
\begin{eqnarray}
\mathbf{E}\; &\rightarrow&\; \mathbf{E}' = \gamma(\mathbf{E}+\boldsymbol{\beta}\times\mathbf{B}) -\frac{\gamma^2}{\gamma+1}\boldsymbol{\beta}(\boldsymbol{\beta}\cdot\mathbf{E}),\\
\mathbf{B}\; &\rightarrow&\; \mathbf{B}' = \gamma(\mathbf{B}-\boldsymbol{\beta}\times\mathbf{E}) -\frac{\gamma^2}{\gamma+1}\boldsymbol{\beta}(\boldsymbol{\beta}\cdot\mathbf{B}),\qquad
\end{eqnarray}
\end{subequations}
and $\pi^\alpha=\left(c^{-1}(H-e\phi),\boldsymbol{\pi}\right)$ transforms as a 4-vector:
\begin{subequations}\label{Lorentz trans of pi}
\begin{eqnarray}
\label{Lorentz trans of pi a}
H-e\phi\; &\rightarrow&\; H'-e\phi' = \gamma\left(H-e\phi-c\,\boldsymbol{\beta}\cdot\boldsymbol{\pi}\right),\qquad\\
\label{Lorentz trans of pi b}
\boldsymbol{\pi}\; &\rightarrow&\; \boldsymbol{\pi}' = \boldsymbol{\pi} +\frac{\gamma-1}{\boldsymbol{\beta}^2}(\boldsymbol{\beta}\cdot\boldsymbol{\pi})\boldsymbol{\beta}\nonumber\\ &&\qquad\quad-\frac{\gamma}{c}\boldsymbol{\beta}(H-e\phi),
\end{eqnarray}
\end{subequations}
where $\boldsymbol{\beta}:=\mathbf{v}/c$, $\gamma:=(1-\boldsymbol{\beta}^2)^{-1/2}$,\footnote{\label{foot:gamma factors}Do not confuse $\gamma$ with $\gamma_{\boldsymbol{\pi}}$ in \eqnref{T-BMT new F}. The former is the Lorentz factor due to reference boost, while the latter is the Lorentz factor associated with the particle's momentum $\boldsymbol{\pi}$.} and
\begin{eqnarray}
H'&:=& H(\mathbf{p}',\phi',\mathbf{A}',\mathbf{E}',\mathbf{B}')\\
&=&H'_\mathrm{orbit}+H'_\mathrm{spin}+\mathcal{O}(F_{\mu\nu}^2,\hbar^2)\nonumber\\
&=& H_\mathrm{orbit}(\boldsymbol{\pi}',\phi') + \mathbf{s}\cdot \mathbf{F}_{\boldsymbol{\pi}}(\boldsymbol{\pi}',\mathbf{E}',\mathbf{B}') +\mathcal{O}(F_{\mu\nu}^2,\hbar^2).\nonumber
\end{eqnarray}
In the weak-field limit, $H_\mathrm{spin}\ll mc^2$ as discussed in \secref{sec:Hamiltonian formalism} and can be neglected on the right-hand side of \eqnref{Lorentz trans of pi b}; consequently, \eqnref{Lorentz trans of pi a} leads to
\begin{equation}
\mathbf{F}_{\boldsymbol{\pi}}(\boldsymbol{\pi}',\mathbf{E}',\mathbf{B}') =\gamma\mathbf{F}_{\boldsymbol{\pi}}(\boldsymbol{\pi},\mathbf{E},\mathbf{B})
+\mathcal{O}(F_{\mu\nu}^2,\hbar^2).
\end{equation}
That is, if we apply the replacement rules \eqnref{Lorentz trans of E B} and \eqnref{Lorentz trans of pi b} (with $H_\mathrm{spin}$ part neglected) to $\mathbf{F}_{\boldsymbol{\pi}}$ given in \eqnref{T-BMT new F}, all occurrences of $\boldsymbol{\beta}$ and $\gamma$ shall be intricately balanced (up to corrections of $\mathcal{O}(F_{\mu\nu}^2,\hbar^2)$) to yield a single overall $\gamma$ factor.
It is however very difficult to explicitly see the intricate balance take place as it essentially relies on the very involved derivations we have performed to obtain \eqnref{T-BMT4}--\eqnref{T-BMT F}.

As an illustrative example, in particular, let the boost velocity $\mathbf{v}$ be parallel to $\boldsymbol{\pi}$. A moment of reflection tells that
\begin{equation}\label{particular boost}
\mathbf{v}\equiv c\,\boldsymbol{\beta} =(\cdots)\boldsymbol{\pi},
\qquad
\gamma=(\cdots),
\end{equation}
and, with the $H_\mathrm{spin}$ part neglected,
\begin{equation}\label{Lorentz trans of pi 2}
\boldsymbol{\pi}\; \rightarrow\; \boldsymbol{\pi}'=(\cdots)\boldsymbol{\pi},
\end{equation}
where $(\cdots)$ denotes some functions of $\boldsymbol{\pi}^2$.
The transformation rule \eqnref{Lorentz trans of E B} for this case then reads as
\begin{subequations}\label{Lorentz trans of E B 2}
\begin{eqnarray}
\mathbf{E}\; &\rightarrow&\; \mathbf{E}' = (\cdots)\mathbf{E} +(\cdots)\boldsymbol{\pi}\times\mathbf{B}\nonumber\\ &&\qquad\quad\mbox{}+(\cdots)\boldsymbol{\pi}(\boldsymbol{\pi}\cdot\mathbf{E}),\\
\mathbf{B}\; &\rightarrow&\; \mathbf{B}' = (\cdots)\mathbf{B} +(\cdots)\boldsymbol{\pi}\times\mathbf{E}\nonumber\\ &&\qquad\quad\mbox{}+(\cdots)\boldsymbol{\pi}(\boldsymbol{\pi}\cdot\mathbf{B}).
\end{eqnarray}
\end{subequations}
Applying \eqnref{Lorentz trans of pi 2} and \eqnref{Lorentz trans of E B 2} to \eqnref{T-BMT new F} and considering $\mathbf{B}'$, $(\boldsymbol{\pi}'\cdot\mathbf{B}')\boldsymbol{\pi}'$ and $\boldsymbol{\pi}'\times\mathbf{E}'$, we have
\begin{subequations}\label{match up}
\begin{eqnarray}
\label{match up a}
(\cdots)\mathbf{B}' &=& (\cdots)\mathbf{B} +(\cdots)\boldsymbol{\pi}\times\mathbf{E}\nonumber\\
&&\mbox{} +(\cdots)(\boldsymbol{\pi}\cdot\mathbf{B})\boldsymbol{\pi}\\
\label{match up b}
(\cdots)(\boldsymbol{\pi}'\cdot\mathbf{B}')\boldsymbol{\pi}' &=& (\cdots)(\boldsymbol{\pi}\cdot\mathbf{B})\boldsymbol{\pi} +(\cdots)\left[\boldsymbol{\pi}\cdot(\boldsymbol{\pi}\times\mathbf{E})\right]\boldsymbol{\pi}\nonumber\\
&=& (\cdots)(\boldsymbol{\pi}\cdot\mathbf{B})\boldsymbol{\pi} +(\cdots)\left[(\boldsymbol{\pi}\times\boldsymbol{\pi})\cdot\mathbf{E}\right]\boldsymbol{\pi}\nonumber\\
&=& (\cdots)(\boldsymbol{\pi}\cdot\mathbf{B})\boldsymbol{\pi},\\
\label{match up c}
(\cdots)(\boldsymbol{\pi}'\times\mathbf{E}') &=& (\cdots)\boldsymbol{\pi}\times\mathbf{E} +(\cdots)\boldsymbol{\pi}\times(\boldsymbol{\pi}\times\mathbf{B})\nonumber\\
&&\mbox{} +(\cdots)(\boldsymbol{\pi}\times\boldsymbol{\pi})(\boldsymbol{\pi}\cdot\mathbf{E})\nonumber\\
&=& (\cdots)\boldsymbol{\pi}\times\mathbf{E} +(\cdots)\mathbf{B}\nonumber\\ &&\mbox{}+(\cdots)(\boldsymbol{\pi}\cdot\mathbf{B})\boldsymbol{\pi}.
\end{eqnarray}
\end{subequations}
Apart from $(\cdots)$ factors, the terms appearing on the right-hand sides are $\mathbf{B}$, $(\boldsymbol{\pi}\cdot\mathbf{B})\boldsymbol{\pi}$ and $\boldsymbol{\pi}\times\mathbf{E}$. It thus demonstrates that, under the Lorentz boost at least for the case of \eqnref{particular boost}, \eqnref{T-BMT new F} remains in the same form (multiplied by an overall $\gamma$ factor), on the assumption that those $(\cdots)$ conspire to yield the desired functions of $\boldsymbol{\pi}^2$ (through $\gamma_{\boldsymbol{\pi}}$).

The fact that $\mathbf{F}_{\boldsymbol{\pi}}$ is form invariant (apart from the overall $\gamma$ factor) should have a correspondence for the relativistic quantum theory. That is, if we formally change all ``unprimed'' variables to ``primed'' ones in \eqnref{H total FW} and then \emph{formally} apply the replacement rules \eqnref{Lorentz trans of pi 2} and \eqnref{Lorentz trans of E B 2}, the $H_\mathrm{spin}$ part of the FW transformed Hamiltonian in \eqnref{H total FW} should end up in the same form (multiplied by $\gamma$) up to $\mathcal{O}(F_{\mu\nu}^2,\hbar^2)$.\footnote{Note that \eqnref{particular boost} does not make sense at the quantum level, as the right-hand side is an operator while the left-hand side is a parameter (also recall \footref{foot:gamma factors}). Nevertheless, it is still expected that \emph{formally} applying the replacement rules \eqnref{Lorentz trans of pi 2} and \eqnref{Lorentz trans of E B 2} should yield the result in accord with the classical one, since \eqnref{particular boost} regarded as assigning a specific value for $\mathbf{v}$ does make sense at the classical level as far as equalities in \eqnref{Lorentz trans of pi} are concerned.} However, as variables are operators in quantum theory, further complications have to be taken into consideration. First, $\boldsymbol{\pi}\times\boldsymbol{\pi}$ no longer vanishes. Fortunately, $\boldsymbol{\pi}\times\boldsymbol{\pi}$ is of $\mathcal{O}(F_{\mu\nu})$ and thus $[(\boldsymbol{\pi}\times\boldsymbol{\pi})\cdot\mathbf{E}]\boldsymbol{\pi}$ and $(\boldsymbol{\pi}\times\boldsymbol{\pi})(\boldsymbol{\pi}\cdot\mathbf{E})$ are negligible  up to $\mathcal{O}(F_{\mu\nu})$ when we try to reproduce \eqnref{match up b} and \eqnref{match up c}.
Second, \eqnref{H total FW} involves a specific way of operator ordering and this causes the main complication.

At the heuristic level, let us consider only the orderings \eqnref{pi dot F pi} and \eqnref{pi times F} but disregard the Weyl ordering over powers of $\boldsymbol{\pi}^2$ as specified in \eqnref{Weyl ordering}, assuming that all terms involving powers of $\boldsymbol{\pi}^2$ conspire to yield the desired operator ordering. Therefore, instead of $(\boldsymbol{\pi}'\cdot\mathbf{B}')\boldsymbol{\pi}'$ and $\boldsymbol{\pi}'\times\mathbf{E}'$ in \eqnref{match up}, we should consider $\overline{(\boldsymbol{\pi}'\cdot\mathbf{B}')\boldsymbol{\pi}'}$ and $\overline{\boldsymbol{\pi}'\times\mathbf{E}'}$. As $\overline{\boldsymbol{\pi}'\times\mathbf{E}'}$ gives rise to $\overline{\boldsymbol{\pi}\times\overline{\boldsymbol{\pi}\times\mathbf{B}}}$, in order to reproduce the match-up as in \eqnref{match up c}, we must have $\overline{\boldsymbol{\pi}\times\overline{\boldsymbol{\pi}\times\mathbf{B}}} \sim \overline{(\boldsymbol{\pi}\cdot\mathbf{B})\boldsymbol{\pi}} -\boldsymbol{\pi}^2\mathbf{B}$ as a necessary condition, where $\sim$ denotes equality up to the ordering over powers of $\boldsymbol{\pi}^2$. If we adopt the operator ordering in \eqnref{pi times F}, we then have
\begin{eqnarray}
\overline{\boldsymbol{\pi}\times\overline{\boldsymbol{\pi}\times\mathbf{B}}}
&=&\frac{1}{4}\Big(
\boldsymbol{\pi}\times(\boldsymbol{\pi}\times\mathbf{B})
-\boldsymbol{\pi}\times(\mathbf{B}\times\boldsymbol{\pi})\\
&&\quad\mbox{}
-(\boldsymbol{\pi}\times\mathbf{B})\times\boldsymbol{\pi}
+(\mathbf{B}\times\boldsymbol{\pi})\times\boldsymbol{\pi}
\Big),\qquad\nonumber
\end{eqnarray}
or equivalently,
\begin{eqnarray}
&&4\left(\overline{\boldsymbol{\pi}\times\overline{\boldsymbol{\pi}\times\mathbf{B}}}\right)_i
= \epsilon_{ijk}\epsilon_{klm}\left[\pi_j\pi_lB_m-\pi_jB_l\pi_m\right]\nonumber\\
&& \qquad\qquad\qquad\qquad
-\epsilon_{ijk}\epsilon_{jlm}\left[\pi_lB_m\pi_k-B_l\pi_m\pi_k\right]\nonumber\\
&=&\pi_j\pi_iB_j-\pi_j\pi_jB_i-\pi_jB_i\pi_j+\pi_jB_j\pi_i\nonumber\\
&&\mbox{}-\pi_jB_i\pi_j+\pi_iB_j\pi_j+B_j\pi_i\pi_j-B_i\pi_j\pi_j,
\end{eqnarray}
where $\epsilon_{kij}\epsilon_{klm}=\delta_{il}\delta_{jm}-\delta_{im}\delta_{jl}$ has been used. Noting that $[\pi_i,\pi_j]$ is of $\mathcal{O}(F_{\mu\nu})$ and disregarding the ordering over $\boldsymbol{\pi}^2\equiv\pi_j\pi_j$, we then have
\begin{eqnarray}
\overline{\boldsymbol{\pi}\times\overline{\boldsymbol{\pi}\times\mathbf{B}}}
&\sim&
\frac{1}{4}\Big(
\boldsymbol{\pi}(\boldsymbol{\pi}\cdot\mathbf{B})
+(\boldsymbol{\pi}\cdot\mathbf{B})\boldsymbol{\pi}\\
&&\quad\mbox{}
+\boldsymbol{\pi}(\mathbf{B}\cdot\boldsymbol{\pi})
+(\mathbf{B}\cdot\boldsymbol{\pi})\boldsymbol{\pi}
\Big)
-\boldsymbol{\pi}^2\mathbf{B}.\nonumber
\end{eqnarray}
The condition that the right-hand side has to take the form of $\overline{(\boldsymbol{\pi}\cdot\mathbf{B})\boldsymbol{\pi}} -\boldsymbol{\pi}^2\mathbf{B}$ dictates that the ordering for $\overline{(\boldsymbol{\pi}\cdot\mathbf{F})\boldsymbol{\pi}}$ must be the one prescribed by \eqnref{pi dot F pi}.
This fixes the operator ordering for the second term inside the square brackets in \eqnref{H total FW}.

We should keep in mind that the argument above is very heuristic and issues concerning ordering ambiguity still demand further investigations. In quantum mechanics, the ordering ambiguity is a long-standing problem and quite relevant in some physically important systems (see \cite{Shewell1959} for a review).
Notably, in the study of semiconductor heterostructures (and, more generally, of inhomogeneous crystals), the model with a position-dependent effective mass arising from the envelope-function approximation has generated extensive discussions on the issues of operator ordering (see \cite{Roos1983,Levy-Leblond1995} and references therein).

It was argued that Hamiltonians with position-dependent masses are not Galilean invariant and in general there exist many nonequivalent Hamiltonians within the same envelope-function approximation \cite{Roos1983}. On the other hand, from a more fundamental point of view, it was shown that the use of position-dependent effective masses gives correct approximations and indeed is a conceptually consistent approach, whereby considerations of instantaneous Galilean invariance lead to a specific family of acceptable Hamiltonians and fix the ordering ambiguity (as well as the boundary conditions at abrupt interfaces) \cite{Levy-Leblond1995}.
Furthermore, the relationship between the ordering ambiguity and exact solvability of the Schr\"{o}dinger equation with position-dependent masses has been discussed in \cite{Dutra2000}; the exactly solvable PT-symmetric potentials of the Dirac equation in 1+1 dimensions with position-dependent masses were also presented in \cite{Jia2008}.

There is a close analogy between the FW transformed Dirac-Pauli equation subject to inhomogeneous electromagnetic fields for charged spinors and the Schr\"{o}dinger equation with position-dependent effective masses for semiconductor heterostructures. The Dirac-Pauli equation, which is self-consistent only in the context of quantum field theory, is analogous to the many-body equation, which describes the fundamental physics of a position-dependent composition; the block-diagonalized wavefunction via the FW transformation in the low-energy/weak-field limit is analogous to the one-electron envelope function in the envelope-function approximation, with inhomogeneous electromagnetic fields analogous to position-dependent masses; and Lorentz invariance is analogous to Galilean invariance. The analogy to the model of semiconductor heterostructures might offer valuable insight about operator orderings of the FW transformed Hamiltonian.

The Dirac-Pauli equation does not have any ordering ambiguity; after the FW transformation, it dictates a unique operator ordering in the FW transformed Hamiltonian as specified by \eqnref{pi dot F pi}--\eqnref{Weyl ordering}. The uniqueness of ordering seems to be a consequence of the fact that the FW transformation in the low-energy/weak-field limit is consistent with the Lorentz invariance, by analogy to the model of semiconductor heterostructures. Even though we only found the ordering without ambiguity for two special cases, the ordering should be extended to generic cases, as suggested in \cite{Dutra2000} for the exact solvability of potentials with position-dependent masses.
However, it is still possible that other orderings can be obtained for different cases or by different FW scenarios, since different orderings could turn out to be equivalent as demonstrated in \cite{Dutra2000} for operators with linear dependence on the momentum.
Further research is needed to better understand the operator ordering and its relevance of the FW transformed Dirac-Pauli Hamiltonian.

\section{Summary and discussion}\label{sec:summary}
In the presence of inhomogeneity of electromagnetic fields, the BMT equation for a relativistic classical spinor is modified to \eqnref{BMT} with the extra term involving $f^\alpha$, which accounts for any forces on the orbital motion other than the Lorentz force such as the field gradient (e.g., Stern-Gerlach) force as denoted in \eqnref{covariant Lorentz eq}. It turns out, whether $f^\alpha$ is taken into account or not, the corresponding T-BMT equation remains unaltered, as given in \eqnref{T-BMT5} with \eqnref{T-BMT F}.

To describe the orbital motion and the spin precession simultaneously, in the Hamiltonian formalism with $\mathbf{x}$, $\mathbf{p}$, and $\mathbf{s}$ as independent phase-space variables, the total Hamiltonian is given by \eqnref{H total modified} with \eqnref{H spin modified} and \eqnref{T-BMT new F}. The spin Hamiltonian $H_\mathrm{spin}(\mathbf{s},\mathbf{x},\mathbf{p};t)$ is given by the prescription which replaces every occurrence of the velocity $\mathbf{v}$ in \eqnref{H spin} and \eqnref{T-BMT F} with the ``velocity'' $\mathbf{v}_{\boldsymbol{\pi}}$ associated with the kinematic momentum $\boldsymbol{\pi}$, defined in \eqnref{v pi}.

The prescription of replacing $\mathbf{v}$ with $\mathbf{v}_{\boldsymbol{\pi}}$ is justifiable, as the BMT and T-BMT equations are accordingly altered if the constraint on the spin 4-vector is imposed by \eqnref{constraint 1 modified} instead of \eqnref{constraint 1}. As the FW transformed Dirac-Pauli Hamiltonian corresponds to the prescribed classical Hamiltonian, it seems to suggest that the Pauli matrices $\sigma_i$ used in the Dirac-Pauli equation represent the intrinsic spin vector (via $\mathbf{s}=\hbar\boldsymbol{\sigma}/2$) for the ``rest frame'' comoving with $\mathbf{v}_{\boldsymbol{\pi}}$, instead of $\mathbf{v}$. The difference is, however, inappreciable in the weak-field limit, and further investigation on the strong-field ramifications of the Dirac-Pauli equation is needed to tell whether it is indeed the case.

Meanwhile, close examination of Lorentz covariance reveals that the classical theory described by \eqnref{H total modified} exactly preserves Lorentz invariance only if the negligible $\mathcal{O}(F_{\mu\nu}^2,\hbar^2)$ terms are appropriately supplemented by a more fundamental theory such as the Dirac-Pauli equation, which indeed yields (infinitely many) nonlinear electromagnetic terms of $\mathcal{O}(F_{\mu\nu}^2)$ as well as the Darwin term of $\mathcal{O}(\hbar^2)$. It is natural to ask: without referring to the relativistic quantum theory, can we construct a fundamental \emph{classical} theory of a spinor which is explicitly Lorentz invariant and gives Hamiltonian in the form of \eqnref{H total modified} in the weak-field limit? Although the Lorentz-invariant Lagrangian formalism for a classical relativistic point particle endowed with intrinsic spin has been formulated in various approaches (see $\S$II of \cite{Barut1980}, \cite{Halbwachs:1960}, and references therein), none of them seems to give the same weak-field description as that by \eqnref{H total modified} and thus the question remains an open problem.

For the relativistic quantum theory, we consider two special cases of which the interaction term is an odd
Dirac matrix, and apply Eriksen's method to obtain the exact FW transformation of the Dirac-Pauli Hamiltonian for both cases. These two cases are complementary to each other, as the first case is with $\mathbf{E}=0$ and $\mu'=0$ (zero anomalous magnetic moment) while the second with $\mathbf{B}=0$ and $e=0$ (zero charge). In the low-energy/weak-field limit in which nonlinear terms in electromagnetic fields are neglected, these two cases in conjunctions strongly suggest that the FW transformed Dirac-Pauli Hamiltonian is given in the form of \eqnref{H total FW}, which is in full agreement with the classical Hamiltonian given by \eqnref{H total modified}--\eqnref{T-BMT new F} (with $\mathbf{s}=\hbar\boldsymbol{\sigma}/2$) plus the Darwin term multiplied by the $1/\gamma_{\boldsymbol{\pi}}$ factor. In this correspondence, classical variables in the classical Hamiltonian is promoted to quantum operators via the specific Weyl ordering defined in \eqnref{pi dot F pi}--\eqnref{Weyl ordering}. Issues regarding operator orderings remain a matter of further investigation; particularly, considerable insight might be obtained by exploiting the analogy to the model of semiconductor heterostructures.

The FW transformation reveals that the magnetic moment of a Dirac-Pauli spinor is given by \eqnref{magnetic moment}, where $\mu'$ gives rise to the anomalous magnetic moment. When $e$ and $\mu'$ are of the same sign, the magnitude of the magnetic moment is increased by the presence of $\mu'$, but the Darwin term, by contrast, is decreased as indicated by the last term in \eqnref{H total FW}. Furthermore, the Darwin term has shown to have no classical correspondence at the level of classical equations of motion.

Throughout this paper, we have focused on the case of inhomogeneous electromagnetic fields. We wonder what will happen if the external fields are both nonstatic and inhomogeneous. For the classical dynamics, the BMT equation shall remain in the same form of \eqnref{BMT}, as the new force (if any) due to time variation of fields can still be attributed to $f^\alpha$ in \eqnref{covariant Lorentz eq}. Consequently, the T-BMT equation remains the same as \eqnref{T-BMT5}. For the relativistic quantum dynamics, on the other hand, it is unclear whether the Darwin term has a counterpart involving $\partial_t\mathbf{E}$ and/or $\partial_t\mathbf{B}$.

Finally, as the FW transformation has been extended to the spin-$1$ particles \cite{Case1954,Silenko2004,Silenko2013} and even generalized to the case of arbitrary spins \cite{Jayaraman1975,Silenko2009}, it is natural and important to ask whether the correspondence we observed between the classical and Dirac-Pauli spinors can be extended to the case of arbitrary spins. We anticipate the answer to be affirmative, as the magnitude of $\mathbf{s}$ is arbitrary in the classical dynamics.

\vspace{.25in}

\begin{acknowledgments}
T.W.C.\ was supported by the National Science Council of Taiwan under Contract No.\ NSC 101-2112-M-110-013-MY3; D.W.C.\ was supported by the Center for Advanced Study in Theoretical Sciences at National Taiwan University.
\end{acknowledgments}

\end{document}